\documentclass{moriond}

\usepackage{siunitx}
\usepackage[italic]{hepnames}
\usepackage{xspace}
\usepackage[capitalise]{cleveref}
\usepackage{bm}
\usepackage{hyperref}

\usepackage{lineno}

\def\be{\begin{equation}}
\def\ee{\end{equation}}
\def\bea{\begin{eqnarray}}
\def\eea{\end{eqnarray}}

\newcommand{\Photo}{}

\newcommand{\absVub}{\ensuremath{|V_{\Pup\Pbottom}|}\xspace}

\newcommand{\absVcb}{\ensuremath{|V_{\Pcharm\Pbottom}|}\xspace}

\newcommand{\PX}{\HepParticle{X}{}{}\xspace}

\newcommand{\PXc}{\HepParticle{X}{c}{}\xspace}

\newcommand{\PKstzero}{\HepParticle{K}{}{*0}\xspace}
\newcommand{\PKstplus}{\HepParticle{K}{}{*+}\xspace}

\newcommand{\PDbostbc}{\HepParticle{D}{}{(*)}\xspace}

\newcommand{\PDstminus}{\HepParticle{D}{}{*-}\xspace}

\newcommand{\BtoXlv}{\HepProcess{\PB \to \PX \Plepton \Pnulepton}\xspace}
\newcommand{\BtoXclv}{\HepProcess{\PB \to \PXc \Plepton \Pnulepton}\xspace}

\begin{document}
\vspace*{4cm}
\title{Latest results on semileptonic and electroweak penguin decays at Belle~II}

\author{M. WELSCH on behalf of the Belle~II Collaboration }

\address{Physikalisches Institut, Nu\ss{}allee~12\\
University of Bonn, 53115 Bonn, Germany}

\maketitle

\abstracts{The Belle~II Collaboration presents four new analyses: The measurement of \absVub from \HepProcess{\PB \to \Ppi \Pe \Pnue} decays with a fit to the differential $q^2$ spectrum, the determination of \absVcb using a fit to the differential $w$ distribution from \HepProcess{\PBzero \to \PDstminus \Pleptonplus \Pnulepton} decays, and the measurement of $q^2$ moments in inclusive \BtoXclv decays.
In all these analyses tag-side \PB meson is reconstructed in a fully hadronic decay chain. 
The last result are branching fraction measurement of \HepProcess{\PB \to \PKst \Pleptonplus \Pleptonminus} decays using an untagged approach.}

\section{Introduction}
The Belle~II Collaboration collected a data set corresponding to an integrated $\SI{189}{fb^{-1}}$ at the \PUpsilonFourS resonance. 
The data is recorded with the Belle~II detector and the SuperKEKB $\Ppositron\Pelectron$-collider located at KEK in Tsukuba, Japan. 
This manuscript presents four new recent results from the Belle~II Collaboration.

\section{Branching Fraction Measurements of \HepProcess{\PB \to \Ppi \Ppositron \Pnue} Decays and Determination of \absVub} 

In this analysis reconstructs the decays \HepProcess{\PBzero \to \Ppiminus \Ppositron \Pnue} and \HepProcess{\PBplus\to \Ppizero \Ppositron \Pnue} using hadronic tagging provided by tagging algorithm implemented in the Full-Event-Interpretation ~\cite{Keck:2018lcd}.
These decays are considered the golden modes to measure the magnitude of the Cabibbo-Kobayashi-Maskawa (CKM)~\cite{PhysRevLett.10.531,km_paper} matrix element \absVub in an exclusive approach.


\HepProcess{\PB \to \Ppi \Ppositron \Pnue} decays are reconstructed in channels with charged and neutral \Ppi candidates.
The number of signal and background events are determined with a maximum likelihood fit of the missing mass squared $m_\mathrm{Miss}^2 = (p_{\PB_\mathrm{sig}}^\ast - p^\ast_{\Pe} - p^\ast_{\Ppi})^2 $ in three bins of the four-momentum transfer $q^2 =  (p_{\PB_\mathrm{sig}}^\ast - p^\ast_{\Ppi})^2$.
Here, $p^\ast$ denotes the four momentum in the center-of-mass of the $\Ppositron\Pelectron$-collision.
The four-momentum of the signal \PB is calculated from the momentum of fully reconstructed tag-side \PB meson $
    p_{\PB_\mathrm{sig}}^\ast = (\frac{m_{\PUpsilonFourS}}{2}, -\bm{p}^\ast_{\PB_\mathrm{tag}})$.

\begin{figure}[b]
    \centering
    \includegraphics[width=0.4\textwidth]{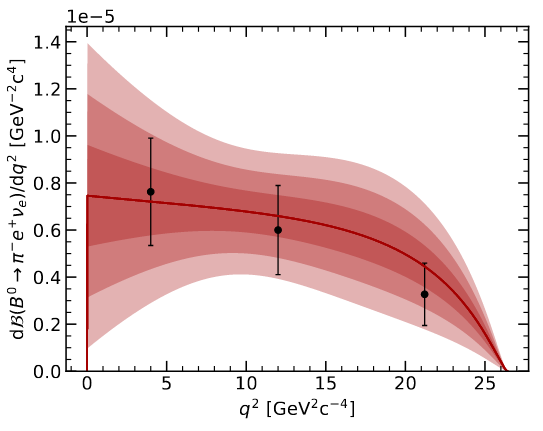}
    \includegraphics[width=0.4\textwidth]{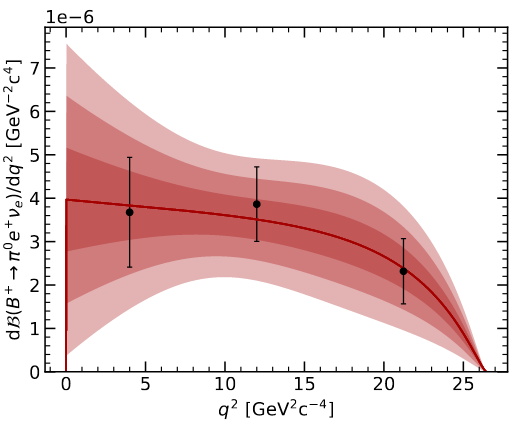}
    \caption{Resulting projections for the charged (left) and neutral (right) pion channel of the combined fit of the differential branching fraction $\mathrm{d}\mathcal{B}/\mathrm{d}q^2$.}
    \label{fig:vub_combined_fit}
\end{figure}

In a next step, the unfolded yields are translated into partial branching fractions in bins of $q^2$.
A $\chi^2$-fit of  $\mathrm{d}\mathcal{B}/\mathrm{d}q^2$ to the measured partial branching fractions and to the Fermilab/MILC lattice QCD constraints~\cite{PhysRevD.92.034506} is performed for each reconstruction  channel individually and and the combination of charged and neutral \PB modes.
In the fit, the form factors are parameterized using the BCL parameterization~\cite{PhysRevD.79.013008}.
The fit result for the combined fit is shown in \cref{fig:vub_combined_fit}.
\Cref{tab:vub_values} lists the extracted \absVub values.
In addition, we report the total branching fractions $\mathcal{B}( \HepProcess{\PBzero  \to \Ppiminus \Ppositron \Pnue}) = (1.43 \pm 0.27_\mathrm{stat} \pm 0.07_\mathrm{sys})\times 10^{-4}$ and $\mathcal{B}( \HepProcess{\PBplus  \to \Ppizero \Ppositron \Pnue}) = (8.43 \pm 1.42_\mathrm{stat} \pm 0.54_\mathrm{sys})\times 10^{-5}$ obtained by the sum of the partial branching fractions in the three $q^2$ bins.

\begin{table}
    \centering
    \caption{Extracted \absVub values by $\chi^2$-fits to $\mathrm{d}\mathcal{B}/\mathrm{d}q^2$ of  $\PBzero  \to \Ppiminus \Ppositron \Pnue$ and $\PBplus  \to \Ppizero \Ppositron \Pnue$ decays.}
    \label{tab:vub_values}
    \vspace{0.4cm}
    \begin{tabular}{lrr}
    \hline
    Decay mode &  $\absVub \times 10^{3}$ &  $\chi^2$/DOF \\
    \hline
    $\PBzero  \to \Ppiminus \Ppositron \Pnue$ & (3.71 $\pm$ 0.55)  & 0.16\\
    $\PBplus  \to \Ppizero \Ppositron \Pnue$ &(4.21 $\pm$ 0.63)  & 0.02 \\
    Combined fit & (3.88 $\pm$ 0.45) & 0.32 \\
    \hline
    \end{tabular}
\end{table}

\section{Branching Fraction Measurements of \HepProcess{\PBzero \to \PDstminus \Pleptonplus \Pnulepton} Decays and Determination of \absVcb}

In this analysis, the decay \HepProcess{\PBzero \to \PDstminus \Pleptonplus \Pnulepton} is reconstructed using a hadronic tagging approach.
A good understanding of the decay \HepProcess{\PB \to \PDst \Plepton \Pnulepton} is important for future measurements of $R(\PDbostbc)$ and the precise exclusive determination of the CKM matrix element \absVcb.

Events are reconstructed in a single decay chain, where the \PDst decays into a neutral \APD meson and a charged pion \Ppiminus. 
The \APD meson is reconstructed in the decay channel \HepProcess{\APD \to  \PKplus \Ppiminus}.

\begin{figure}[bt]
    \centering
    \includegraphics[width=0.4\textwidth]{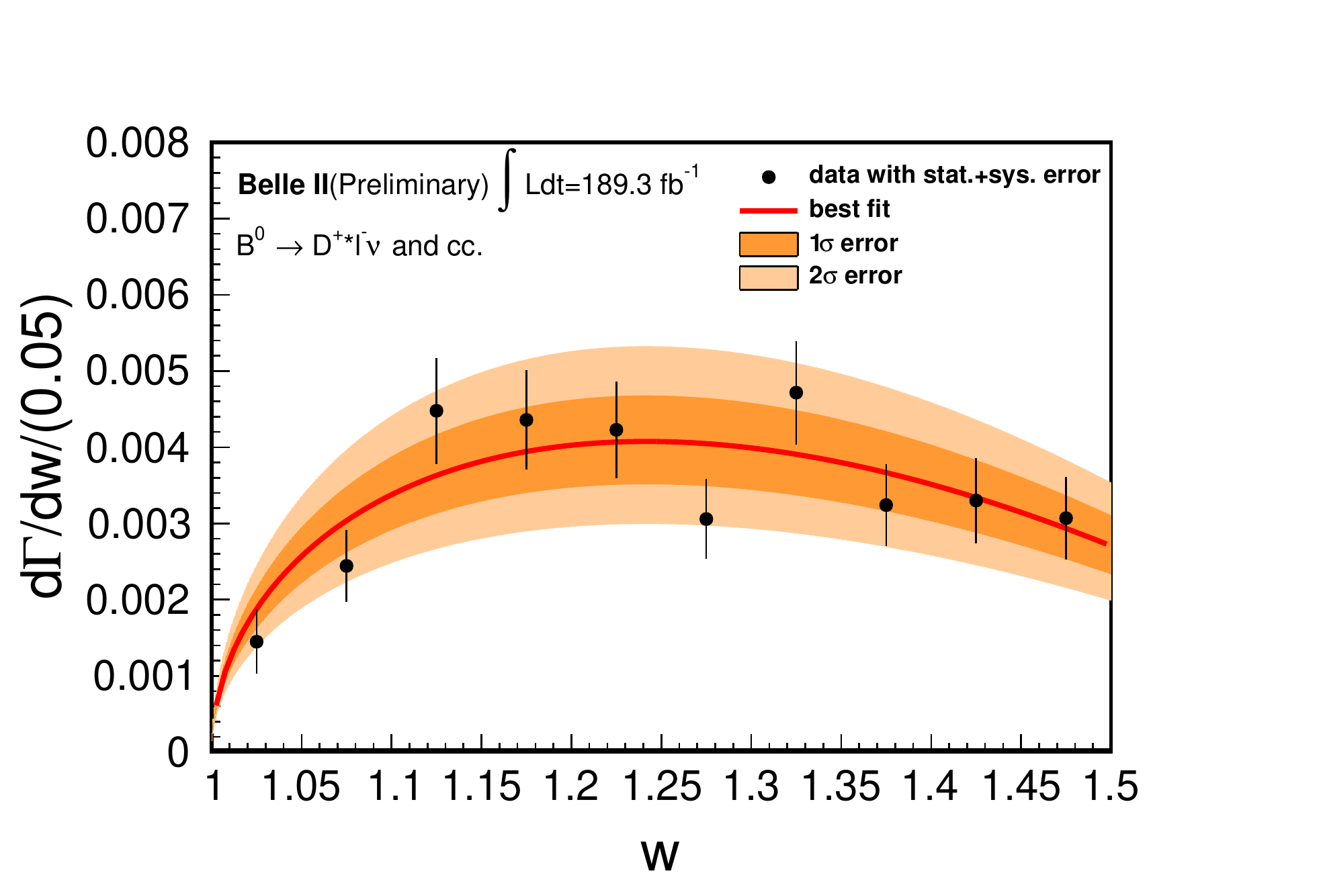}
    \includegraphics[width=0.4\textwidth]{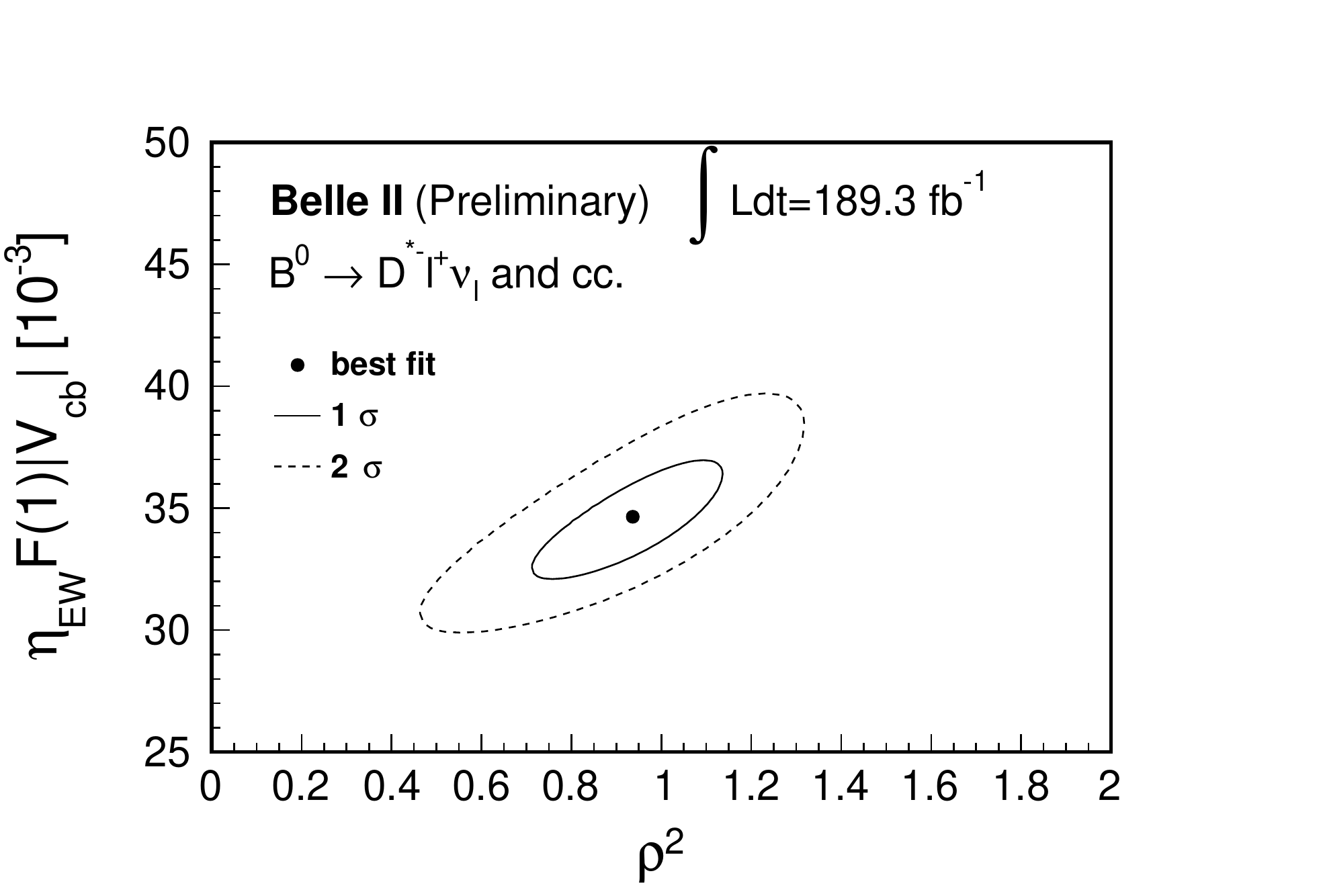}
    \caption{Result of the fit of the differential decay rate $ \mathrm{d}\Gamma / \mathrm{d}w$ (left) and 2D scan of the $\chi^2$ function in the $\rho^2$ and $\eta_\mathrm{EW}F(1)\absVcb$ plane (right). }
    \label{fig:vcb_fit_result}
\end{figure}

We measure a branching fraction of $\mathcal{B}(\HepProcess{\PBzero \to \PDstminus \Pleptonplus \Pnulepton}) = (5.27 \pm 0.22_\mathrm{stat}  \pm 0.38_\mathrm{sys} ) \% $.
\absVcb~is determined with a fit of the differential decay rate to the unfolded $w$ spectrum.
Here, $w$ denotes the product of \PB and \PDst four-velocities $ w = v_{\PB} v_{\PDst} = (m_{\PB}^2 + m_{\PDst}^2 - q^2)/(2m_{\PB}m_{\PDst})$.
The differential decay rate is given by
\begin{linenomath*}
\begin{align}
    \frac{\mathrm{d}\Gamma}{\mathrm{d}w} = \frac{\eta_\mathrm{EW}^2 G_\mathrm{F}^3}{48\pi^2} m_{\PDst}^3( m_{\PB} - m_{\PDst})^2 g(w) F^2(w) \absVcb^2.
\end{align}
\end{linenomath*}
The product $g(w) F^2(w)$ is parameterized with $R_1(1)$, $R_2(1)$ and $\rho^2$ in the CLN form factor parameterization~\cite{Caprini_1998}.
The result of a $\chi^2$ fit of $ \mathrm{d}\Gamma / \mathrm{d}w$ is shown in \cref{fig:vcb_fit_result}.
$R_1(1)$ and $R_2(1)$ are constrained to external measurements~\cite{Amhis_2021} in the fit.
A 2D scan of the $\chi^2$ function in the $\rho^2$ and $\eta_\mathrm{EW}F(1)\absVcb$ plane is also shown in \cref*{fig:vcb_fit_result}.
With $\eta_\mathrm{EW} = 1.0066$~\cite{PhysRevD.89.114504} and $F(1)=0.906\pm0.004_\mathrm{stat}\pm 0.0012_\mathrm{sys}$~\cite{PhysRevD.89.114504} we obtain $\absVcb = (37.9 \pm 2.7)\times 10^{-3}$.

\section{Measurement of $q^2$ Moments in \BtoXclv Decays}
We present measurements of the first to fourth raw and central $q^2$ moments of \BtoXclv decays as functions of lower $q^2$ threshold from $\SI{1.5}{GeV^2/c^4}$ up to $\SI{8.5}{GeV^2/c^4}$.
Lepton energy and hadronic mass moments have been used to determine \absVcb and parameters of the Heavy-Quark-Expansion (HQE) up to $\mathcal{O}(1/m_{\Pbottom}^3)$ in a model independent way using a global fit.
At higher order, the determination of the HQE is complicated by a proliferation of non-perturbative matrix elements.
Ref.~\cite{Fael:2018vsp} outlines an alternative approach to determine \absVcb from inclusive decays avoiding this proliferation of HQE parameters  by exploiting reparameterization invariance.
However reparameterization invariance is not retained by the lepton energy and hadronic mass moments.
Thus, the measurement of $q^2$ moments is proposed.

\begin{figure}[bt]
    \centering
    \includegraphics[width=0.4\textwidth]{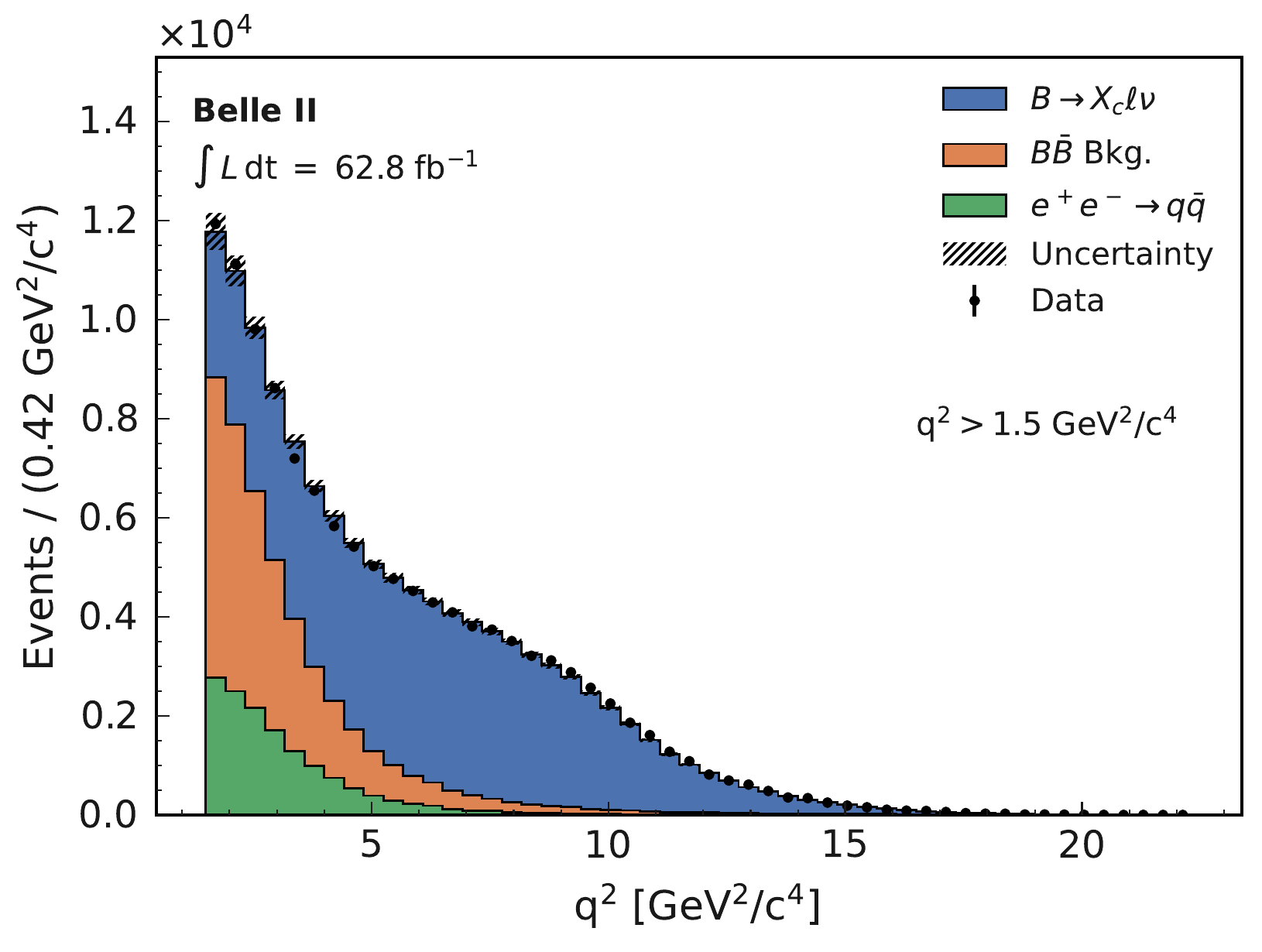}
    \includegraphics[width=0.4\textwidth]{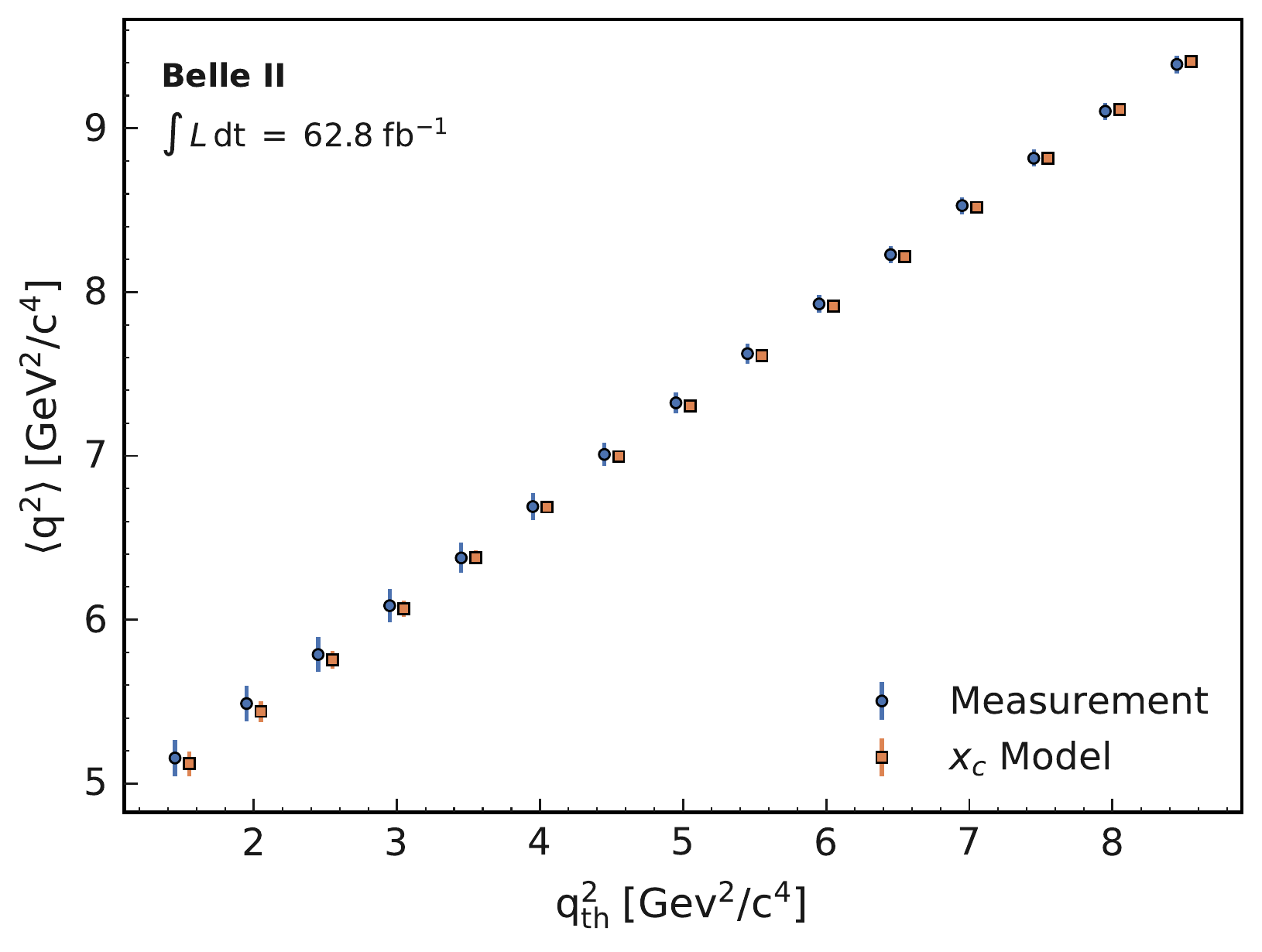}
    \caption{Reconstructed $q^2$ spectrum (left) and first raw $q^2$ moment $\langle q^2 \rangle$ (right).}
    \label{fig:q2_moments}
\end{figure}

Semileptonic decays are reconstructed in an inclusive approach without assuming exclusive hadronic charmed resonances in the the \BtoXlv decay.
The hadronic \PX system is identified with the remaining charged particles and photons not used in the tag-side \PB meson and signal lepton reconstruction.
\Cref{fig:q2_moments} shows the reconstructed $q^2$ spectrum.

Before calculating the moments, a calibration of the $q^2$ spectrum $q^2_\mathrm{reco} \rightarrow q^2_\mathrm{calib}$ is applied to correct for detector resolution and reconstruction efficiency effects.
The remaining background is subtracted directly in the calculation of the $q^2$ moments by assigning a signal probability as event weight $w_i(q^2)$.
The raw moments are calculated as a weighted mean
\begin{linenomath*}
\begin{align}
    \langle q^{2n} \rangle = \frac{\sum w_i q^2_{i,\mathrm{calib}}}{\sum w_j} \times \mathcal{C}_\mathrm{calib} \times \mathcal{C}_\mathrm{gen}.
\end{align}
\end{linenomath*}
Here, $\mathcal{C}_\mathrm{calib}$ and $\mathcal{C}_\mathrm{gen}$ are two additional correction factors correcting imperfections of the calibration procedure and for different selection efficiencies of the \BtoXclv components.
The measurement of the first raw moment $\langle q^2 \rangle$ are also shown in \cref*{fig:q2_moments}.
The moments for the assumed \PXc model are show for comparison.
We provide numerical results and covariance matrices on \href{https://www.hepdata.net}{HEPData}.


\section{Branching Fraction Measurements of \HepProcess{\PB \to \PKst \Pleptonplus \Pleptonminus} Decays}
The decay \HepProcess{\PB \to \PKst \Pleptonplus \Pleptonminus} involves a \HepProcess{\Pbottom \to \Pstrange} transition mediated by a flavour changing neutral current which is forbidden at tree level and thus, highly suppressed.
The decay is sensitive to new physics contributions, which might enhance or suppress the decay amplitude or modify the angular distributions.
The here reported branching fractions measurements are a first step towards a measurement of $R(\PKst)$ at Belle~II.
Recent determinations of $R(\PKst)$ show tensions with the SM expectation~\cite{Aaij_2017,Aaij_2020,PhysRevLett.126.161801}.

The analysis considers three decay modes: \HepProcess{\PBzero \to  \PKstzero (\to \PKplus \Ppiminus) \Pleptonplus \Pleptonminus} and \HepProcess{\PBplus \to  \PKstplus (\to \PKshort \Ppiplus, \to \PKplus \Ppizero) \Pleptonplus \Pleptonminus}.
The second \PB meson in the \PUpsilonFourS decay is not explicitly reconstructed.
The number of signal decays is extracted from a 2D maximum likelihood fit to the beam-constrained mass $M_\mathrm{bc} = \sqrt{ s/4 - p_{\PB}^{\ast 2}}$ and the energy difference $\Delta E = (E_{\PB}^\ast - \sqrt{s}/2$).

We measure $\mathcal{B}(\HepProcess{\PB \to \PKst  \APmuon \Pmuon }) = (1.28 \pm 0.29 ^{+0.08}_{-0.07})  \times 10^{-6}$,  $\mathcal{B}(\HepProcess{\PB \to \PKst \Ppositron \Pelectron}) =  (1.04 \pm 0.48 ^{+0.09}_{-0.09}) \times 10^{-6}$ and $\mathcal{B}(\HepProcess{\PB \to \PKst \Pleptonplus \Pleptonminus}) = (1.22 \pm 0.28 ^{+0.08}_{-0.07}) \times 10^{-6}$ with the first and second uncertainty being statistical and systematic, respectively.


\section*{References}
\bibliography{welsch.bib}






\end{document}